\documentclass[10pt,twocolumn,letterpaper]{article}

\usepackage{cvpr}
\usepackage{times}
\usepackage{epsfig}
\usepackage{graphicx}
\usepackage{amsmath}
\usepackage{amssymb}

\usepackage[utf8]{inputenc} 
\usepackage[T1]{fontenc}    
\usepackage{url}            
\usepackage{booktabs}       
\usepackage{amsfonts}       
\usepackage{amsmath, amsthm, amssymb}
\usepackage{nicefrac}       
\usepackage{microtype}      
\usepackage{xspace}
\usepackage{enumerate}
\usepackage{graphicx}
\usepackage{gensymb}
\usepackage{subfigure}
\usepackage{bbm}
\usepackage{caption}
\usepackage{cite}
\usepackage{array}
\usepackage{mathtools}
\usepackage{authblk}

\newcommand{\Attack}[1]{\mathcal{A}\left(#1\right)}
\usepackage{color}

\newif\ifsubmit
\newif\ifinception

\submittrue
\inceptiontrue

\ifinception
\newcommand{\incep}[1]{#1}
\else
\newcommand{\incep}[1]{}
\fi

\ifsubmit
\newcommand{\dawn}[1]{}
\newcommand{\bo}[1]{}
\newcommand{\kevin}[1]{}
\newcommand{\ef}[1]{}
\newcommand{\ivan}[1]{}
\newcommand{\amir}[1]{}
\newcommand{\yoshi}[1]{}
\newcommand{\wh}[1]{}
\else
\newcommand{\dawn}[1]{\textcolor{red}{Dawn: #1}}
\newcommand{\bo}[1]{\textcolor{blue}{Bo: #1}}
\newcommand{\kevin}[1]{\textcolor{cyan}{Kevin: #1}}
\newcommand{\ef}[1]{\textcolor{red}{(Earl: #1)}}
\newcommand{\ivan}[1]{\textcolor{green}{Ivan: #1}}
\newcommand{\amir}[1]{\textcolor{cyan}{Amir: #1}}
\newcommand{\yoshi}[1]{\textcolor{blue}{Yoshi: #1}}
\newcommand{\wh}[1]{\textcolor{red}{wh: #1}}
\fi

\newcommand{\algfull}[0]{Robust Physical Perturbations\xspace}
\newcommand{\algshort}[0]{RP\textsubscript{2}\xspace}
\newcommand{\classBattackupper}[0]{Camouflage\xspace}
\newcommand{\classBattacklower}[0]{camouflage\xspace}
\newcommand{\pertAupper}[0]{Subtle\xspace}
\newcommand{\pertAlower}[0]{subtle\xspace}

\newcommand{\ignore}[1]{}
\newcommand{\eat}[1]{}

\newcommand{\shallownet}[0]{LISA-CNN\xspace}
\newcommand{\deepnet}[0]{GTSRB-CNN\xspace}

\newcommand{\webpage}[0]{https://iotsecurity.eecs.umich.edu/\#roadsigns}\xspace

\theoremstyle{plain}

\newcolumntype{M}[1]{>{\centering\arraybackslash}m{#1}}
   
\newcommand{\stickerArtSuccessRate}[0]{100\%\xspace} 
\newcommand{\stickerGraffitiSuccessRate}[0]{66.67\%\xspace} 
\newcommand{\invisibleSuccessRate}[0]{100\%\xspace} 
\newcommand{\rightTurnSuccessRate}[0]{73.33\%\xspace} 

\newcommand{\stickerArtSuccessRateDriveByShallowCNN}[0]{84.8\%\xspace} 
\newcommand{\subtlePosterSuccesRateDriveByShallowCNN}[0]{100\%\xspace} 
\newcommand{\stickerArtSuccessRateDriveByDeepCNN}[0]{87.5\%\xspace} 
\newcommand{\stickerArtSuccessRateDeepCNNControlled}[0]{80\%\xspace} 

\cvprfinalcopy 

\def\cvprPaperID{3407} 

\title{Robust Physical-World Attacks on Deep Learning Visual Classification}

\author[1]{Kevin Eykholt\thanks{These authors contributed equally.}}
\author[2]{Ivan Evtimov\textsuperscript{*}}
\author[2]{Earlence Fernandes}
\author[3]{Bo Li}
\author[4]{\\ Amir Rahmati}
\author[1]{Chaowei Xiao}
\author[1]{Atul Prakash}
\author[2]{Tadayoshi Kohno}
\author[3]{Dawn Song}
\affil[1]{University of Michigan, Ann Arbor}
\affil[2]{University of Washington}
\affil[3]{University of California, Berkeley}
\affil[4]{Samsung Research America and Stony Brook University}

\usepackage{fancyhdr}

\begin{document}



\maketitle
\thispagestyle{fancy}

\begin{abstract}
Recent studies show that the state-of-the-art deep neural networks (DNNs) are vulnerable to adversarial examples, resulting from small-magnitude perturbations added to the input. Given that that emerging physical systems are using DNNs in safety-critical situations, adversarial examples could mislead these systems and cause dangerous situations. Therefore, understanding adversarial examples in the physical world is an important step towards developing resilient learning algorithms. We propose a general attack algorithm, \algfull (\algshort), to generate robust {\em visual} adversarial perturbations under different physical conditions. Using the real-world case of road sign classification, we show that adversarial examples generated using \algshort achieve high targeted misclassification rates against standard-architecture road sign classifiers in the physical world under various environmental conditions, including viewpoints. Due to the current lack of a standardized testing method, we propose a two-stage evaluation methodology for robust physical adversarial examples consisting of lab and field tests. Using this methodology, we evaluate the efficacy of physical adversarial manipulations on real objects. With a perturbation in the form of only black and white stickers, we attack a real stop sign, causing targeted misclassification in \stickerArtSuccessRate of the images obtained in lab settings, and in \stickerArtSuccessRateDriveByShallowCNN of the captured video frames obtained on a moving vehicle (field test) for the target classifier. 

\end{abstract}

\section{Introduction}
Deep Neural Networks (DNNs) have achieved state-of-the-art, and sometimes human-competitive, performance on many computer vision tasks~\cite{krizhevsky2012imagenet,taigman2014deepface,Le:2011:LHI:2191740.2192108}. Based on these successes, they are increasingly being used as part of control pipelines in physical systems such as cars~\cite{lillicrap2015continuous,geiger2012we}, UAVs~\cite{bou2010controller,mostegel2016uav}, and robots~\cite{zhang2015towards}.
Recent work, however, has demonstrated that DNNs are vulnerable to adversarial perturbations~\cite{goodfellow2014explaining,li2014feature,li2015scalable,szegedy2014intriguing,papernot2016limitations,carlini2017towards,sabour2015adversarial,kos2017adversarial,universalap,nguyen2015deep}. These carefully crafted modifications to the (visual) input of DNNs can cause the systems they control to misbehave in unexpected and potentially dangerous ways.  


This threat has gained recent attention, and work in computer vision has made great progress in understanding the space of adversarial examples, beginning in the digital domain (\eg by modifying images corresponding to a scene)~\cite{universalap,nguyen2015deep,szegedy2014intriguing,goodfellow2014explaining}, and more recently in the physical domain~\cite{sharif2016accessorize,kurakin2016adversarial,openaiwork,athalye2017synthesizing}. Along similar lines, our work contributes to the understanding of adversarial examples when perturbations are physically added to the \textit{objects themselves}. We choose road sign classification as our target domain for several reasons: (1) The relative visual simplicity of road signs make it challenging to hide perturbations. (2) Road signs exist in a noisy unconstrained environment with changing physical conditions such as the distance and angle of the viewing camera, implying that physical adversarial perturbations should be robust against considerable environmental instability. (3) Road signs play an important role in transportation safety. (4) A reasonable threat model for transportation is that an attacker might not have control over a vehicle's systems, but is able to modify the objects in the physical world that a vehicle might depend on to make crucial safety decisions.

The main challenge with generating robust physical perturbations is environmental variability. Cyber-physical systems operate in noisy physical environments that can destroy perturbations created using current digital-only algorithms~\cite{noneed}. For our chosen application area, the most dynamic environmental change is the distance and angle of the viewing camera. Additionally, other practicality challenges exist: (1) Perturbations in the digital world can be so small in magnitude that it is likely that a camera will not be able to perceive them due to sensor imperfections. (2) Current algorithms produce perturbations that occupy the background imagery of an object. It is extremely difficult to create a robust attack with background modifications because a real object can have varying backgrounds depending on the viewpoint. (3) The fabrication process (e.g., printing of perturbations) is imperfect.

\begin{figure}[t]
  \centering
  \includegraphics[width=1\columnwidth]{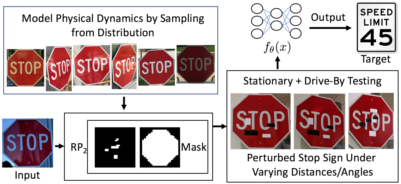}
  \caption{\algshort pipeline overview. The input is the target Stop sign. \algshort samples from a distribution that models physical dynamics (in this case, varying distances and angles), and uses a mask to project computed perturbations to a shape that resembles graffiti. The adversary prints out the resulting perturbations and sticks them to the target Stop sign.}
  \label{fig:pipeline}
\end{figure}

Informed by the challenges above, we design \emph{\algfull (\algshort)}, which can generate perturbations robust to widely changing distances and angles of the viewing camera. \algshort creates a visible, but inconspicuous perturbation that only perturbs the object (\eg a road sign) and not the object's environment. To create robust perturbations, the algorithm draws samples from a distribution that models physical dynamics (\eg varying distances and angles) using experimental data and synthetic transformations (Figure~\ref{fig:pipeline}). 

\begin{figure}[t]
  \centering
  \includegraphics[height=3cm]{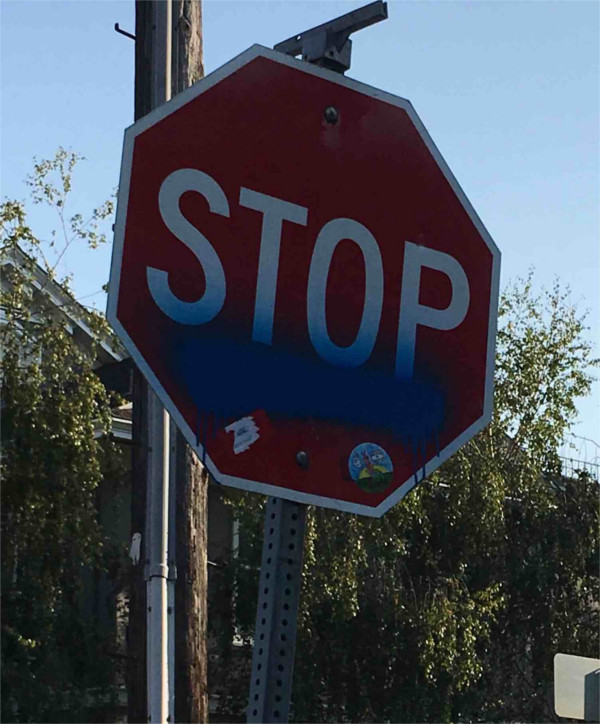}
  \includegraphics[height=3cm]{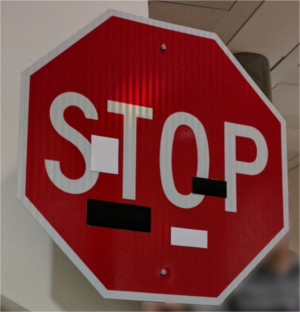}
  \caption{The left image shows real graffiti on a Stop sign, something that most humans would not think is suspicious. The right image shows our a physical perturbation applied to a Stop sign. We design our perturbations to mimic graffiti, and thus ``hide in the human psyche.''}
  \label{fig:real_world}
\end{figure}

Using the proposed algorithm, we evaluate the effectiveness of perturbations on physical objects, and show that adversaries can physically modify objects using low-cost techniques to reliably cause classification errors in DNN-based classifiers under widely varying distances and angles. For example, our attacks cause a classifier to interpret a subtly-modified physical Stop sign as a Speed Limit 45 sign. Specifically, our final form of perturbation is a set of black and white stickers that an adversary can attach to a physical road sign (Stop sign). We designed our perturbations to resemble graffiti, a relatively common form of vandalism. It is common to see road signs with random graffiti or color alterations in the real world as shown in Figure~\ref{fig:real_world} (the left image is of a real sign in a city). If these random patterns were adversarial perturbations (right side of Figure~\ref{fig:real_world} shows our example perturbation), they could lead to severe consequences for autonomous driving systems, without arousing suspicion in human operators.



Given the lack of a standardized method for evaluating physical attacks, we draw on standard techniques from the physical sciences and propose a two-stage experiment design: (1) A lab test where the viewing camera is kept at various distance/angle configurations; and (2) A field test where we drive a car towards an intersection in uncontrolled conditions to simulate an autonomous vehicle. We test our attack algorithm using this evaluation pipeline and find that the perturbations are robust to a variety of distances and angles.

\noindent\textbf{Our Contributions.} Figure~\ref{fig:pipeline} shows an overview of our pipeline to generate and evaluate robust physical adversarial perturbations.
\begin{enumerate}
    \item We introduce \algfull (\algshort) to generate physical perturbations for \textit{physical-world} objects that can consistently cause misclassification in a DNN-based classifier under a range of dynamic physical conditions, including different viewpoint angles and distances (Section~\ref{sec:algo}).
    
    \item Given the lack of a standardized methodology in evaluating physical adversarial perturbations, we propose an evaluation methodology to study the effectiveness of physical perturbations in real world scenarios
    (Section~\ref{sec:methodology}). 
    
    \item  We evaluate our attacks against two standard-architecture classifiers that we built: \shallownet with 91\% accuracy on the LISA test set and \deepnet with 95.7\% accuracy on the GTSRB test set. Using two types of attacks (object-constrained poster and sticker attacks) that we introduce, we show that \algshort produces robust perturbations for real road signs. For example, poster attacks are successful in 100\% of stationary and drive-by tests against \shallownet, and sticker attacks are successful in  \stickerArtSuccessRateDeepCNNControlled of stationary testing conditions and in \stickerArtSuccessRateDriveByDeepCNN of the extracted video frames against \deepnet.
    
    \incep{\item To show the generality of our approach, we generate the robust physical adversarial example by manipulating general physical objects, such as a microwave. We show that the pre-trained Inception-v3 classifier misclassifies the microwave as ``phone" by adding a single sticker.}
    
\end{enumerate}

Our work, thus, contributes to understanding the susceptibility of image classifiers to robust adversarial modifications of \textit{physical objects}.
These results provide a case for the potential consequences of adversarial examples on deep learning models that interact with the physical world through vision. Our overarching goal with this work is to inform research in building robust vision models and to raise awareness on the risks that future physical learning systems might face. 
We include more examples and videos of the drive-by tests on our webpage \webpage

\section{Related Work}
\label{sec:related}
We survey the related work in generating adversarial examples. Specifically, 
given a classifier $f_\theta(\cdot)$ with parameters $\theta$ and an input $x$ with ground truth label $y$ for $x$, an adversarial example $x'$ is generated so that it is close to $x$ in terms of certain distance, such as $L_p$ norm distance. $x'$ will also cause the classifier to make an incorrect prediction as $f_\theta(x') \ne y$ (untargeted attacks), or $f_\theta(x') = y^*$ (targeted attacks) for a specific $y^* \not= y$. We also discuss recent efforts at understanding the space of physical adversarial examples.

\noindent\textbf{Digital Adversarial Examples.} 
Different methods have been proposed to generate adversarial examples in the white-box setting, where the adversary has full access to the classifier~\cite{szegedy2014intriguing,goodfellow2014explaining,carlini2017towards,moosavi2015deepfool,papernot2016limitations,biggio2013evasion,kurakin2016adversarial}. We focus on the white-box setting as well for two reasons: (1) In our chosen autonomous vehicle domain, an attacker can obtain a close approximation of the model by reverse engineering the vehicle's systems using model extraction attacks~\cite{tramer2016stealing}. (2) To develop a foundation for future defenses, we must assess the abilities of powerful adversaries, and this can be done in a white-box setting. Given that recent work has examined the black-box transferability of digital adversarial examples~\cite{papernot2016transferability}, physical black-box attacks may also be possible.



Goodfellow \etal proposed the fast gradient method that applies a first-order approximation of the loss function to construct adversarial samples~\cite{goodfellow2014explaining}. 
Optimization based methods have also been proposed to create adversarial perturbations for targeted attacks~\cite{carlini2017towards,liu2016delving}. These methods contribute to understanding digital adversarial examples. By contrast, our work examines physical perturbations on real objects under varying environmental conditions.

\noindent\textbf{Physical Adversarial Examples.}
Kurakin \etal showed that printed adversarial examples can be misclassified when viewed through a smartphone camera~\cite{kurakin2016adversarial}. 
Athalye and Sutskever improved upon the work of Kurakin \etal and presented an attack algorithm that produces adversarial examples robust to a set of two-dimensional synthetic transformations~\cite{openaiwork}. These works do not modify physical objects---an adversary prints out a digitally-perturbed image on paper. However, there is value in studying the effectiveness of such attacks when subject to environmental variability.  Our object-constrained poster printing attack is a reproduced version of this type of attack, with the additional physical-world constraint of confining perturbations to the surface area of the sign. Additionally, our work goes further and examines how to effectively create adversarial examples where the object itself is physically perturbed by placing stickers on it. 


Concurrent to our work,\footnote{This work appeared at arXiv on 30 Oct 2017.} Athalye \etal improved upon their original attack, and created 3D-printed replicas of perturbed objects~\cite{athalye2017synthesizing}. The main intellectual differences include: (1) Athalye \etal\xspace {\em only} use a set of synthetic transformations during optimization, which can miss subtle physical effects, while our work samples from a distribution modeling both physical {\em and} synthetic transformations. (2) Our work modifies \textit{existing} true-sized objects. Athalye \etal 3D-print small-scale replicas. (3) Our work simulates realistic testing conditions appropriate to the use-case at hand. 

Sharif \etal attacked face recognition systems by printing adversarial perturbations on the frames of eyeglasses~\cite{sharif2016accessorize}. Their work demonstrated successful physical attacks in relatively stable physical conditions with little variation in pose, distance/angle from the camera, and lighting. This contributes an interesting understanding of physical examples in stable environments. However, environmental conditions can vary widely in general and can contribute to reducing the effectiveness of perturbations. Therefore, we choose the inherently unconstrained environment of road-sign classification. In our work, we explicitly design our perturbations to be effective in the presence of diverse physical-world conditions (specifically, large distances/angles and resolution changes).


Finally, Lu \etal performed experiments with physical adversarial examples of road sign images against \textit{detectors} and show current detectors cannot be attacked~\cite{noneed}. In this work, we focus on \textit{classifiers} to demonstrate the physical attack effectiveness and to highlight their security vulnerability in the real world. Attacking detectors are out of the scope of this paper, though recent work has generated digital adversarial examples against detection/segmentation algorithms~\cite{xie2017adversarial,cisse2017houdini,metzen2017universal}, and our recent work has extended \algshort to attack the YOLO detector~\cite{yoloblog}.

\section{Adversarial Examples for Physical Objects}
\label{sec:algo}
Our goal is to examine whether it is possible to create robust physical perturbations for real-world objects that mislead classifiers to make incorrect predictions even when images are taken in a range of varying physical conditions. We first present an analysis of environmental conditions that physical learning systems might encounter, and then present our algorithm to generate physical adversarial perturbations taking these challenges into account.

\subsection{Physical World Challenges}
\label{sec:challenges}
Physical attacks on an object must be able to survive changing conditions and remain effective at fooling the classifier. We structure our discussion of these conditions  
around the chosen example of road sign classification, which could be potentially applied in autonomous vehicles and other safety sensitive domains. A subset of these conditions can also be applied to other types of physical learning systems such as drones, and robots. 

    \noindent\textbf{Environmental Conditions.} The distance and angle of a camera in an autonomous vehicle with respect to a road sign varies continuously. The resulting images that are fed into a classifier are taken at different distances and angles. Therefore, any perturbation that an attacker physically adds to a road sign must be able to survive these transformations of the image. Other environmental factors include changes in lighting/weather conditions, and the presence of debris on the camera or on the road sign. 
    
    
    \noindent\textbf{Spatial Constraints.} Current algorithms focusing on digital images add adversarial perturbations to all parts of the image, including background imagery. However, for a physical road sign, the attacker cannot manipulate background imagery. Furthermore, the attacker cannot count on there being a fixed background imagery as it will change depending on the distance and angle of the viewing camera.
    
    \noindent\textbf{Physical Limits on Imperceptibility.} An attractive feature of current adversarial deep learning algorithms is that their perturbations to a digital image are often so small in magnitude that they are almost imperceptible to the casual observer. However, when transferring such minute perturbations to the real world, we must ensure that a camera is able to perceive the perturbations. Therefore, there are physical limits on how imperceptible perturbations can be, and is dependent on the sensing hardware.
    
    \noindent\textbf{Fabrication Error.} To fabricate the computed perturbation, all perturbation values must be valid colors that can be reproduced in the real world. Furthermore, even if a fabrication device, such as a printer, can produce certain colors, there will be some reproduction error~\cite{sharif2016accessorize}.
    

In order to successfully physically attack deep learning classifiers, an attacker should account for the above categories of physical world variations that can reduce the effectiveness of perturbations.



\subsection{Robust Physical Perturbation} 
We derive our algorithm starting with the optimization method that generates a perturbation for a single image $x$, without considering other physical conditions; then, we describe how to update the algorithm taking the physical challenges above into account.
This single-image optimization problem searches for perturbation $\delta$ to be added to the input $x$, such that the perturbed instance $x' = x + \delta$ is misclassified by the target classifier $f_\theta (\cdot)$:
\begin{align}
\min \quad &H(x+\delta, x),
\quad \text{s.t.} \quad f_\theta(x + \delta) = y^* \nonumber 
\end{align}
where $H$ is a chosen distance function, and $y^*$ is the target class.\footnote{For untargeted attacks, we can modify the objective function to maximize the distance between the model prediction and the true class. We focus on targeted attacks in the rest of the paper.} To solve the above constrained optimization problem efficiently, we reformulate it in the Lagrangian-relaxed form similar to prior work~\cite{liu2016delving,carlini2017towards}.
\begin{equation}
\underset{\delta}{\mathrm{argmin}}~\lambda ||\delta||_{p} + J(f_{\theta}(x + \delta), y^*)
\label{eq:obj-targeted-no-nps}
\end{equation}

Here $J(\cdot~,~\cdot)$ is the loss function, which measures the difference between the model's prediction and the target label $y^*$. $\lambda$ is a hyper-parameter that controls the regularization of the distortion.
We specify the distance function $H$ as $||\delta||_{p}$, denoting the $\ell_{p}$ norm of $\delta$.

Next, we will discuss how the objective function can be modified to account for the \emph{environmental conditions}.
We model the distribution of images containing object $o$ under both physical and digital transformations $X^V$. We
sample different instances $x_i$ drawn from $X^V$.
A physical perturbation can only be added to a specific object $o$ within $x_i$.
In the example of road sign classification, $o$ is the stop sign that we target to manipulate. 
Given images taken in the physical world, we need to make sure that a single perturbation $\delta$, which is added to $o$, can fool the classifier under different physical conditions.
Concurrent work~\cite{athalye2017synthesizing} only applies a set of transformation functions to synthetically sample such a distribution. However, modeling physical phenomena is complex and such synthetic transformations may miss physical effects. Therefore, to better capture the effects of changing physical conditions, we sample instance $x_i$ from $X^V$ by both generating experimental data that contains actual physical condition variability as well as synthetic transformations. For road sign physical conditions, this involves taking images of road signs under various conditions, such as changing distances, angles, and lightning. This approach aims to approximate physical world dynamics more closely.
For synthetic variations, we randomly crop the object within the image, change the brightness, and add spatial transformations to simulate other possible conditions.



To ensure that the perturbations are only applied to the surface area of the target object $o$ (considering the \emph{spatial constraints} and \emph{physical limits on imperceptibility}), we introduce a mask. This mask serves to project the computed perturbations to a physical region on the surface of the object (\ie road sign). In addition to providing spatial locality, the mask also helps generate perturbations that are visible but inconspicuous to human observers. To do this, an attacker can shape the mask to look like graffiti---commonplace vandalism on the street that most humans expect and ignore, therefore hiding the perturbations ``in the human psyche.'' Formally, the perturbation mask is a matrix $M_{x}$ whose dimensions are the same as the size of input to the road sign classifier. $M_{x}$ contains zeroes in regions where no perturbation is added, and ones in regions where the perturbation is added during optimization.

In the course of our experiments, we empirically observed that the position of the mask has an impact on the effectiveness of an attack. We therefore hypothesize that objects have strong and weak physical features from a classification perspective, and we position masks to attack the weak areas. Specifically, we use the following pipeline to discover mask positions: (1) Compute perturbations using the $L_1$ regularization and with a mask that occupies the entire surface area of the sign. $L_1$ makes the optimizer favor a sparse perturbation vector, therefore concentrating the perturbations on regions that are most vulnerable. Visualizing the resulting perturbation provides guidance on mask placement. (2) Recompute perturbations using $L_2$ with a mask positioned on the vulnerable regions identified from the earlier step.

To account for \emph{fabrication error}, we add an additional term to our objective function that models printer color reproduction errors. This term is based upon the Non-Printability Score (NPS) by Sharif \etal~\cite{sharif2016accessorize}. Given a set of printable colors (RGB triples) $P$ and a set $R(\delta)$ of (unique) RGB triples used in the perturbation that need to be printed out in physical world, the non-printability score is given by:
\begin{equation}
\mathit{NPS} = \sum\limits_{\hat{p} \in R(\delta)} \prod \limits_{p' \in P} |\hat{p} - p'|
\label{eq:nps}
\end{equation}


Based on the above discussion, our final robust spatially-constrained perturbation is thus optimized as:
\begin{equation}
\begin{multlined}
\underset{\delta}{\mathrm{argmin}}~\lambda ||M_{x} \cdot \delta||_{p} + \mathit{NPS}\\ +  \mathbb{E}_{x_i \sim X^V} J(f_{\theta}(x_i + T_i( M_{x} \cdot \delta)), y^{*})
\end{multlined}
\label{eq:obj-multi-targeted-nps}
\end{equation}
Here we use function $T_i(\cdot)$ to denote the alignment function that maps transformations on the object to transformations on the perturbation (\eg if the object is rotated, the perturbation is rotated as well). 

Finally, an attacker will print out the optimization result on paper, cut out the perturbation ($M_x$), and put it onto the target object $o$. As our experiments demonstrate in the next section, this kind of perturbation fools the classifier in a variety of viewpoints.\footnote{For our attacks, we use the ADAM optimizer with the following parameters: $\beta_1=0.9$, $\beta_2=0.999$, $\epsilon=10^{-8}$, $\eta \in [10^{-4}, 10^{0}]$}

\section{Experiments}
\label{sec:results}
In this section, we will empirically evaluate the proposed \algshort. We first evaluate a safety sensitive example, Stop sign recognition, to demonstrate the robustness of the proposed physical perturbation. To demonstrate the generality of our approach, we then attack Inception-v3 to misclassify a microwave as a phone.

\subsection{Dataset and Classifiers}
We built two classifiers based on a standard crop-resize-then-classify pipeline for road sign classification as described in~\cite{sermanet2011traffic,papernot2017practical}.
Our \shallownet uses LISA, a U.S. traffic sign dataset containing 47 different road signs~\cite{lisa}. However, the dataset is not well-balanced, resulting is large disparities in representation for different signs. To alleviate this problem, we chose the 17 most common signs based on the number of training examples. 
\shallownet's architecture is defined in the Cleverhans library~\cite{cleverhans} and consists of three convolutional layers and an FC layer. It has an accuracy of 91\% on the test set.

Our second classifier is \deepnet, that is trained on the German Traffic Sign Recognition Benchmark (GTSRB)~\cite{Stallkamp2012}. We use a publicly available implementation~\cite{yadav} of a multi-scale CNN architecture that has been known to perform well on road sign recognition~\cite{sermanet2011traffic}. Because we did not have access to German Stop signs for our physical experiments, we replaced the German Stop signs in the training, validation, and test sets of GTSRB with the U.S. Stop sign images in LISA. \deepnet achieves  95.7\% accuracy on the test set. 
When evaluating \deepnet on our own 181 stop sign images, it achieves 99.4\% accuracy.

\subsection{Experimental Design}
\label{sec:methodology}
To the best of our knowledge, there is currently no standardized methodology of evaluating physical adversarial perturbations. Based on our discussion from Section~\ref{sec:challenges}, we focus on angles and distances because they are the most rapidly changing elements for our use case. A camera in a vehicle approaching a sign will take a series of images at regular intervals. These images will be taken at different angles and distances, therefore changing the amount of detail present in any given image. Any successful physical perturbation must cause targeted misclassification in a range of distances and angles because a vehicle will likely perform voting on a set of frames (images) from a video before issuing a controller action.
Our current experiments do not explicitly control ambient light, and as is evident from experimental data (Section~\ref{sec:results}), lighting varied from indoor lighting to outdoor lighting.

Drawing on standard practice in the physical sciences, our experimental design encapsulates the above physical factors into a two-stage evaluation consisting of controlled lab tests and field tests.



\noindent\textbf{Stationary (Lab) Tests.} This involves classifying images of objects from stationary, fixed positions. 

\begin{enumerate}
    \item Obtain a set of clean images $C$ and a set of adversarially perturbed images ($\{\Attack{c}\}, \forall c\in C$)  at varying distances $d \in D$, and varying angles $g \in G$. We use $c^{d,g}$ here to denote the image taken from distance $d$ and angle $g$. The camera's vertical elevation should be kept approximately constant. 
    Changes in the camera angle relative the the sign will normally occur when the car is turning, changing lanes, or following a curved road.
    
    \item 
    Compute the attack success rate of the physical perturbation using the following formula:  
    \begin{equation}
    \label{eq:lab-success}
    \scalebox{1.5}{$\frac{\sum\limits_{c \in C} \mathbbm{1}_{\{f_{\theta}(\Attack{c^{d,g}}) = y^* \;\wedge f_{\theta}(c^{d,g}) = y\}}}{\sum\limits_{c \in C} \mathbbm{1}_{\{f_{\theta}(c^{d,g}) = y\}}}$}
    \end{equation}
    
    where $d$ and $g$ denote the camera distance and angle for the image, $y$ is the ground truth, and $y^*$ is the targeted attacking class.\footnote{For untargeted adversarial perturbations, change $f_{\theta}(e^{d,g}) = y^*$ to $f_{\theta}(e^{d,g}) \ne y$.} 
\end{enumerate}

Note that an image $\Attack{c}$ that causes misclassification is considered as a successful attack only if the original image $c$ with the same camera distance and angle is correctly classified, which ensures that the misclassification is caused by the added perturbation instead of other factors.

\noindent\textbf{Drive-By (Field) Tests.} We place a camera on a moving platform, and obtain data at realistic driving speeds. For our experiments, we use a smartphone camera mounted on a car.

\begin{enumerate}
    \item Begin recording video at approximately 250 ft away from the sign. Our driving track was straight without curves. Drive toward the sign at normal driving speeds and stop recording once the vehicle passes the sign. In our experiments, our speed varied between 0 mph and 20 mph. This simulates a human driver approaching a sign in a large city. 
    
    \item Perform video recording as above for a ``clean'' sign and for a sign with perturbations applied, and then apply similar formula as Eq.~\ref{eq:lab-success} to calculate the attack success rate, where $C$ here represents the sampled frames.

\end{enumerate}

\begin{table*}[h!]
    \small
    \centering
    \caption{Sample of physical adversarial examples against \shallownet and \deepnet.}
    \resizebox{.87\textwidth}{!}{%
    \begin{tabular}{cM{20mm}M{20mm}M{20mm}M{20mm}M{20mm}M{20mm}}
    \toprule
    Distance/Angle & Subtle Poster & Subtle Poster Right Turn & Camouflage Graffiti & Camouflage Art (\shallownet) & Camouflage Art (\deepnet) \\
   \toprule
    5\texttt{'} 0\degree & \includegraphics[width=0.2\columnwidth]{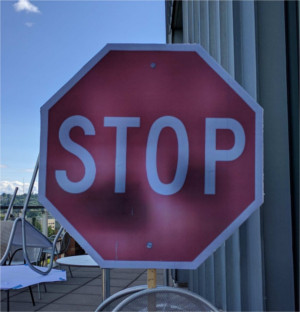} & \includegraphics[width=0.2\columnwidth]{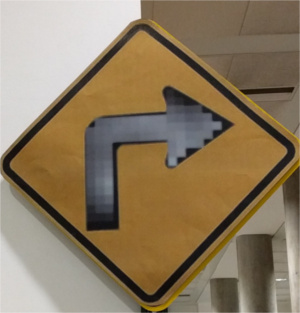} & \includegraphics[width=0.2\columnwidth]{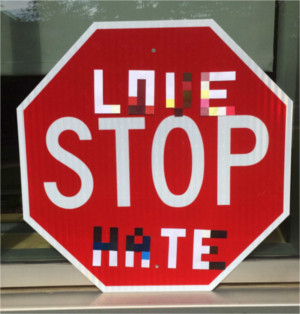} & \includegraphics[width=0.2\columnwidth]{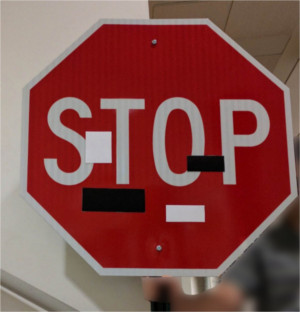} & \includegraphics[width=0.2\columnwidth]{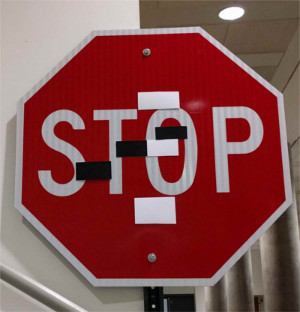} \\ 
    5\texttt{'} 15\degree & \includegraphics[width=0.2\columnwidth]{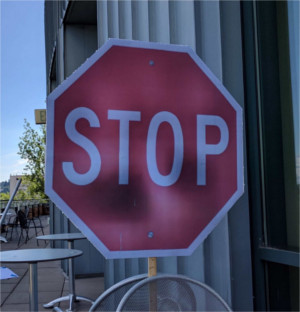} & \includegraphics[width=0.2\columnwidth]{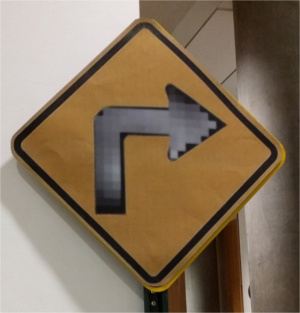} & \includegraphics[width=0.2\columnwidth]{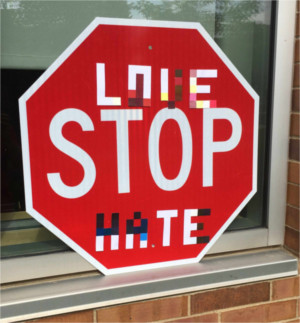} & \includegraphics[width=0.2\columnwidth]{figs/subliminal-art-sticker-05ft-15deg.jpg} & \includegraphics[width=0.2\columnwidth]{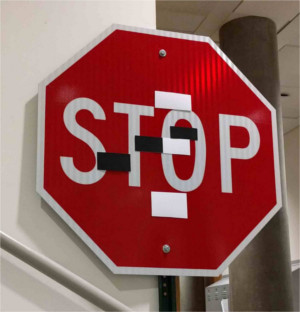} \\ 
    10\texttt{'} 0\degree & \includegraphics[width=0.2\columnwidth]{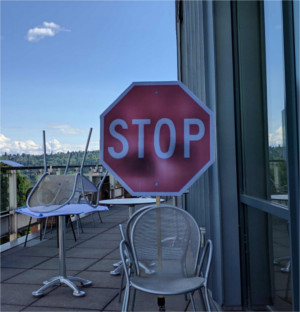} & \includegraphics[width=0.2\columnwidth]{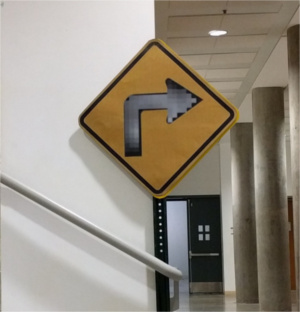} & \includegraphics[width=0.2\columnwidth]{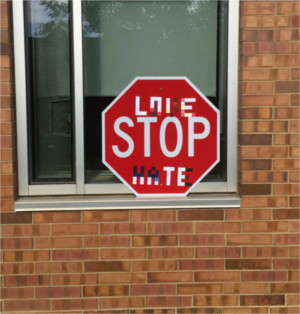} & \includegraphics[width=0.2\columnwidth]{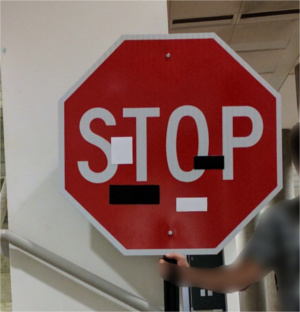} & \includegraphics[width=0.2\columnwidth]{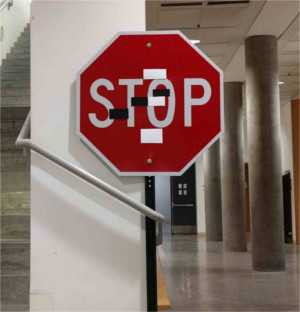} \\ 
    10\texttt{'} 30\degree & \includegraphics[width=0.2\columnwidth]{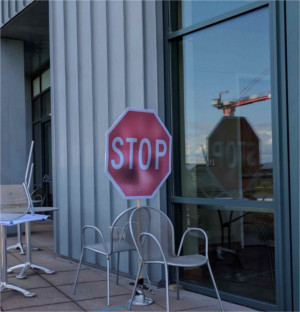} & \includegraphics[width=0.2\columnwidth]{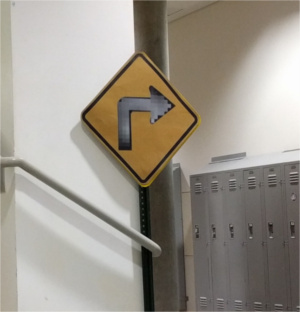} & \includegraphics[width=0.2\columnwidth]{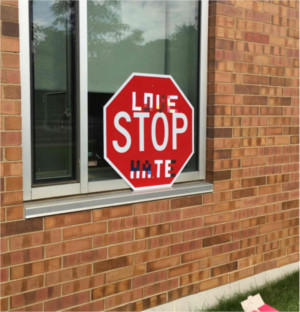} & \includegraphics[width=0.2\columnwidth]{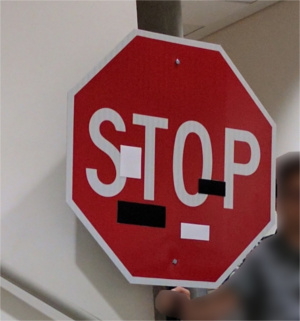} & \includegraphics[width=0.2\columnwidth]{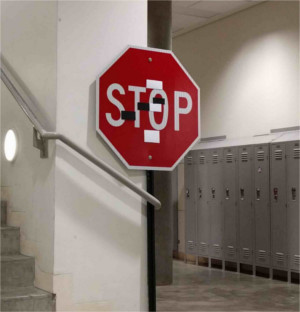} \\
    40\texttt{'} 0\degree & \includegraphics[width=0.2\columnwidth]{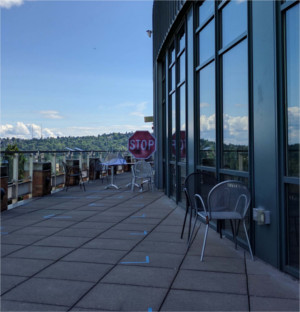} & \includegraphics[width=0.2\columnwidth]{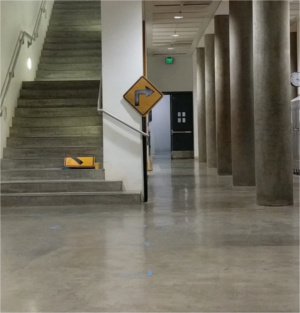} & \includegraphics[width=0.2\columnwidth]{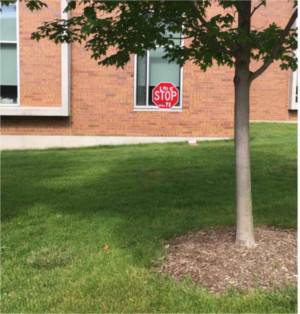} & \includegraphics[width=0.2\columnwidth]{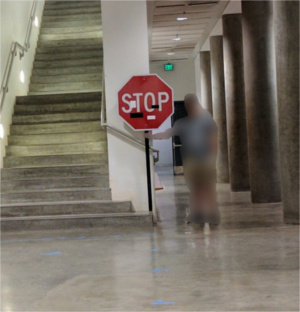} & \includegraphics[width=0.2\columnwidth]{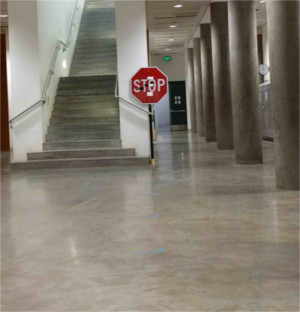} \\ \bottomrule
    Targeted-Attack Success & \invisibleSuccessRate & \rightTurnSuccessRate & \stickerGraffitiSuccessRate & \stickerArtSuccessRate & \stickerArtSuccessRateDeepCNNControlled \\ \bottomrule
    \end{tabular}
    }

\label{tab:sample-experimental-images}
\end{table*}

An autonomous vehicle will likely not run classification on every frame due to performance constraints, but rather, would classify every $j$-th frame, and then perform simple majority voting. Hence, an open question is to determine whether the choice of frame ($j$) affects attack accuracy. In our experiments, we use $j = 10$. We also tried $j = 15$ and did not observe any significant change in the attack success rates.
If both types of tests produce high success rates, the attack is likely to be successful in commonly experienced physical conditions for cars. 

\subsection{Results for \shallownet}

We evaluate the effectiveness of our algorithm by generating three types of adversarial examples on \shallownet (91\% accuracy on test-set). For all types, we observe high attack success rates with high confidence. Table~\ref{tab:sample-experimental-images} summarizes a sampling of stationary attack images. In all testing conditions, our baseline of unperturbed road signs achieves a 100\% classification rate into the true class.

\noindent\textbf{Object-Constrained Poster-Printing Attacks.} This involves reproducing the attack of Kurakin \etal~\cite{kurakin2016adversarial}. The crucial difference is that in our attack, the perturbations are confined to the surface area of the sign excluding the background, and are robust against large angle and distance variations. The Stop sign is misclassified into the attack's target class of Speed Limit 45 in \invisibleSuccessRate of the images taken according to our evaluation methodology. The average confidence of predicting the manipulated sign as the target class is $80.51\%$ (second column of Table~\ref{tab:stop_results}).

\begin{table*}[htb!]
    \small
    \centering
    \caption{Targeted physical perturbation experiment results on \shallownet using a poster-printed Stop sign (\pertAlower attacks) and a real Stop sign (\classBattacklower graffiti attacks, \classBattacklower  art attacks). For each image, the top two labels and their associated confidence values are shown. The misclassification target was Speed Limit 45. See Table~\ref{tab:sample-experimental-images} for example images of each attack.
    Legend: SL45 = Speed Limit 45, STP = Stop, YLD = Yield, ADL = Added Lane, SA = Signal Ahead, LE = Lane Ends.}
    \begin{tabular}{c l l  l l l l}
    \toprule
    Distance \& Angle & \multicolumn{2}{c}{Poster-Printing} & \multicolumn{4}{c}{Sticker} \\ \toprule
                   & \multicolumn{2}{c}{\pertAupper } & \multicolumn{2}{c}{\classBattackupper--Graffiti} & \multicolumn{2}{c}{\classBattackupper--Art} \\ \cmidrule{2-7}
    
    5\texttt{'} 0\degree  & SL45 (0.86)  & ADL (0.03) & \textcolor{red}{STP} (0.40)                    & SL45 (0.27)   & SL45 (0.64)   & LE (0.11) \\
    5\texttt{'} 15\degree & SL45 (0.86)  & ADL (0.02)  & \textcolor{red}{STP} (0.40)   & YLD (0.26)    & SL45 (0.39)   & STP (0.30) \\
    5\texttt{'} 30\degree & SL45 (0.57)  & STP (0.18) & SL45 (0.25)                   & SA (0.18)     & SL45 (0.43)   & STP (0.29)\\
    5\texttt{'} 45\degree & SL45 (0.80)  & STP (0.09) & \textcolor{red}{YLD} (0.21)                    & STP (0.20)    & SL45 (0.37)   & STP (0.31) \\
    5\texttt{'} 60\degree  & SL45 (0.61)  & STP (0.19) & \textcolor{red}{STP} (0.39)   & YLD (0.19)    & SL45 (0.53)   & STP (0.16) \\ \cmidrule{2-7}
    10\texttt{'} 0\degree & SL45 (0.86)  & ADL (0.02) & SL45 (0.48)                   & STP (0.23)    & SL45 (0.77)   & LE (0.04) \\
    10\texttt{'} 15\degree& SL45 (0.90)  & STP (0.02) & SL45	(0.58)                  &	STP (0.21)  & SL45 (0.71)   & STP (0.08) \\
    10\texttt{'} 30\degree& SL45 (0.93)  & STP (0.01) & \textcolor{red}{STP} (0.34)   &	SL45 (0.26) & SL45 (0.47)   & STP (0.30) \\ \cmidrule{2-7}
    15\texttt{'} 0\degree & SL45 (0.81)  & LE (0.05) & SL45 (0.54)                   &  STP (0.22)   & SL45 (0.79)   & STP (0.05) \\
    15\texttt{'} 15\degree & SL45 (0.92)  & ADL (0.01) & SL45 (0.67)                   & STP (0.15)    & SL45 (0.79)   & STP (0.06) \\ \cmidrule{2-7}
    20\texttt{'} 0\degree  & SL45 (0.83)  & ADL (0.03) & SL45 (0.62)                   &	STP (0.18)  & SL45 (0.68)   & STP (0.12) \\
    20\texttt{'} 15\degree& SL45 (0.88)  & STP (0.02) & SL45	(0.70)                  & STP (0.08)    & SL45 (0.67)   & STP (0.11) \\ \cmidrule{2-7}
    25\texttt{'} 0\degree  & SL45 (0.76)  & STP (0.04) & SL45 (0.58)                   &	STP (0.17)  & SL45 (0.67)   & STP (0.08) \\
    30\texttt{'} 0\degree  & SL45 (0.71)  & STP (0.07) & SL45 (0.60)                   &	STP (0.19)  & SL45 (0.76)   & STP (0.10) \\
    40\texttt{'} 0\degree & SL45 (0.78)  & LE (0.04)  & 	SL45 (0.54)                 &	STP (0.21)  & SL45 (0.68)   & STP (0.14) \\
    
    \bottomrule
        
    \end{tabular}
    
    \label{tab:stop_results}
\end{table*}
For the Right Turn warning sign, we choose a mask that covers only the arrow since we intend to generate \pertAlower perturbations. In order to achieve this goal, we increase the regularization parameter $\lambda$ in equation (\ref{eq:obj-multi-targeted-nps}) to demonstrate small magnitude perturbations. Table \ref{tab:turn-right} summarizes our attack results---we achieve a \rightTurnSuccessRate targeted-attack success rate (Table~\ref{tab:sample-experimental-images}). Out of 15 distance/angle configurations, four instances were not classified into the target. However, they were still misclassified into other classes that were not the true label (Yield, Added Lane). Three of these four instances were an Added Lane sign---a different type of warning. We hypothesize that given the similar appearance of warning signs, small perturbations are sufficient to confuse the classifier.

\begin{table*}[tb!]
\small
\center
\caption{Drive-by testing summary for \shallownet. In our baseline test, all frames were correctly classified as a Stop sign. In all attack cases, the perturbations are the same as in Table~\ref{tab:stop_results}. We have added the yellow boxes as a visual guide manually.}
\resizebox{\textwidth}{!}{%
\begin{tabular}{c c c}
\toprule
Perturbation & Attack Success & A Subset of Sampled Frames $k = 10$ \\ \toprule
\pertAupper poster & \subtlePosterSuccesRateDriveByShallowCNN & \includegraphics[width=0.7\textwidth]{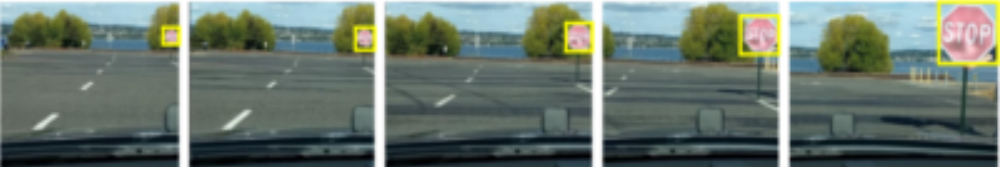} \\ \midrule
\classBattackupper abstract art & \stickerArtSuccessRateDriveByShallowCNN &  \includegraphics[width=0.7\textwidth]{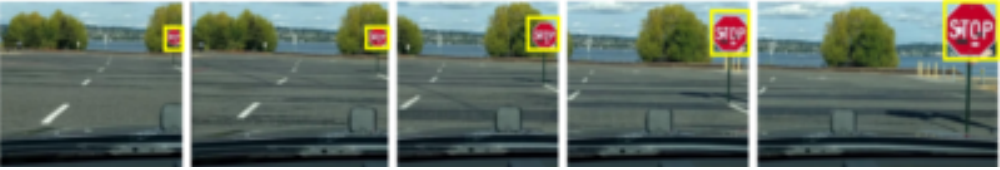} \\ \bottomrule
\end{tabular}
}
\label{tab:driveby91cnn}
\end{table*}

\begin{table}[htbp!]
\small
\center
\caption{Poster-printed perturbation (faded arrow) attack against the \shallownet for a Right Turn sign at varying distances and angles. See example images in Table 1 of the main text. Our targeted-attack success rate is \rightTurnSuccessRate.}
\resizebox{\columnwidth}{!}{%
\begin{tabular}{c l l}
\toprule
Distance \& Angle & Top Class (Confid.) & Second Class (Confid.) \\
\toprule
5' 0\degree &	Stop	(0.39) & Speed Limit 45  (0.10) \\
5' 15\degree &	\textcolor{red}{Yield}	(0.20) & Stop (0.18) \\
5' 30\degree & Stop (0.13) & Yield (0.13) \\
5' 45\degree &	Stop (0.25) &	Yield (0.18) \\
5' 60\degree &	\textcolor{red}{Added Lane} (0.15) & 	Stop  (0.13) \\ \midrule
10' 0\degree &	Stop  (0.29) &  Added Lane  (0.16) \\
10' 15\degree & Stop 	(0.43) &	Added Lane (0.09) \\
10' 30\degree &	\textcolor{red}{Added Lane} (0.19) &	Speed limit 45  (0.16) \\ \midrule
15' 0\degree &	Stop (0.33) &  Added Lane (0.19) \\
15' 15\degree & Stop (0.52) & Right Turn (0.08) \\ \midrule
20' 0\degree &	Stop (0.39) &	Added Lane (0.15) \\
20' 15\degree & Stop	(0.38) & Right Turn (0.11) \\ \midrule
25' 0\degree &	Stop (0.23) &	Added Lane (0.12) \\
30' 0\degree &	Stop (0.23) &	Added Lane (0.15) \\
40' 0\degree &	\textcolor{red}{Added Lane} (0.18) &	 Stop (0.16) \\
\bottomrule
\end{tabular}
}
\label{tab:turn-right}
\end{table}

\noindent\textbf{Sticker Attacks.}
Next, we demonstrate how effective it is to generate physical perturbations in the form of stickers, by constraining the modifications to a region resembling graffiti or art. 
The fourth and fifth columns of Table~\ref{tab:sample-experimental-images} show a sample of images, and Table~\ref{tab:stop_results} (columns 4 and 6) shows detailed success rates with confidences. In the stationary setting, we achieve a \stickerGraffitiSuccessRate targeted-attack success rate for the graffiti sticker attack and a \stickerArtSuccessRate targeted-attack success rate for the sticker \classBattacklower art attack. Some region mismatches may lead to the lower performance of the LOVE-HATE graffiti.

\noindent\textbf{Drive-By Testing.}
Per our evaluation methodology, we conduct drive-by testing for the perturbation of a Stop sign. In our baseline test we record two consecutive videos of a clean Stop sign from a moving vehicle, perform frame grabs at $k = 10$, and crop the sign. We observe that the Stop sign is correctly classified in all frames. We similarly test subtle and abstract art perturbations for \shallownet using $k = 10$. Our attack achieves a targeted-attack success rate of \subtlePosterSuccesRateDriveByShallowCNN for the \pertAlower poster attack, and a targeted-attack success rate of \stickerArtSuccessRateDriveByShallowCNN for the \classBattacklower abstract art attack. See Table \ref{tab:driveby91cnn} for sample frames from the drive-by video.


\begin{table}[tb!]
\small
\center
\caption{A \classBattacklower art attack on \deepnet. See example images in Table~\ref{tab:sample-experimental-images}. The targeted-attack success rate is \stickerArtSuccessRateDeepCNNControlled (true class label: Stop, target: Speed Limit 80).}
\resizebox{\columnwidth}{!}{%
\begin{tabular}{c l l}
\toprule
Distance \& Angle & Top Class (Confid.) & Second Class (Confid.) \\
\toprule

5\texttt{'} 0\degree &	Speed Limit 80 (0.88) & Speed Limit 70  (0.07) \\
5\texttt{'} 15\degree &	Speed Limit 80 (0.94) & Stop (0.03) \\
5\texttt{'} 30\degree &  Speed Limit 80 (0.86) & Keep Right  (0.03) \\
5\texttt{'} 45\degree &	\textcolor{red}{Keep Right} (0.82) &	Speed Limit 80 (0.12) \\
5\texttt{'} 60\degree &	Speed Limit 80 (0.55) & Stop  (0.31) \\ \midrule
10\texttt{'} 0\degree &	Speed Limit 80 (0.98) & Speed Limit 100 (0.006) \\
10\texttt{'} 15\degree & \textcolor{red}{Stop} 	(0.75) & Speed Limit 80 (0.20) \\
10\texttt{'} 30\degree &	Speed Limit 80 (0.77) &	Speed Limit 100  (0.11) \\ \midrule
15\texttt{'} 0\degree &	Speed Limit 80 (0.98) &  Speed Limit 100 (0.01) \\
15\texttt{'} 15\degree & \textcolor{red}{Stop} (0.90) & Speed Limit 80 (0.06) \\ \midrule
20\texttt{'} 0\degree &	Speed Limit 80 (0.95) &	Speed Limit 100 (0.03) \\
20\texttt{'} 15\degree & Speed Limit 80	(0.97) & Speed Limit 100 (0.01) \\ \midrule
25\texttt{'} 0\degree &	Speed Limit 80 (0.99) &	Speed Limit 70 (0.0008) \\
30\texttt{'} 0\degree &	Speed Limit 80 (0.99) &	Speed Limit 100 (0.002) \\
40\texttt{'} 0\degree &	Speed Limit 80 (0.99) &	Speed Limit 100 (0.002) \\
\bottomrule
\end{tabular}
}
\label{tab:95cnnresults}
\end{table}

\subsection{Results for \deepnet}
To show the versatility of our attack algorithms, we create and test attacks for \deepnet (95.7\% accuracy on test-set). Based on our high success rates with the \classBattacklower-art attacks, we create similar abstract art sticker perturbations. The last column of Table~\ref{tab:sample-experimental-images} shows a subset of experimental images. Table~\ref{tab:95cnnresults} summarizes our attack results---our attack fools the classifier into believing that a Stop sign is a Speed Limit 80 sign in \stickerArtSuccessRateDeepCNNControlled of the stationary testing conditions. Per our evaluation methodology, we also conduct a drive-by test ($k = 10$, two consecutive video recordings). The attack fools the classifier \stickerArtSuccessRateDriveByDeepCNN of the time. 

\subsection{Results for Inception-v3}
To demonstrate generality of \algshort, we computed physical perturbations for the standard Inception-v3 classifier~\cite{inceptionv3paper,NIPS2012_4824} using two different objects, a microwave and a coffee mug. We chose a sticker attack since poster printing an entirely new surface for the objects may raise suspicions. Note that for both attacks, we have reduced the range of distances used due to the smaller size of the cup and microwave compared to a road sign (\eg Coffee Mug height-~11.2cm, Microwave height-~24cm, Right Turn sign height-~45cm, Stop Sign-~76cm). Table~\ref{tab:microwave} summarizes our attack results on the microwave and Table~\ref{tab:mug} summarizes our attack results on the coffee mug. For the microwave, the targeted attack success rate is 90\%. For the coffee mug, the targeted attack success rate is 71.4\% and the untargeted success rate is 100\%. Example images of the adversarials stickers for the microwave and cup can be seen in Tables~\ref{tab:microwave-images} and \ref{tab:mug-images}.

\begin{table}[tb!]
\begin{small}
\centering
\caption{Sticker perturbation attack on the Inception-v3 classifier. The original classification is microwave and the attacker's target is phone. See example images in Table~\ref{tab:microwave-images}. Our targeted-attack success rate is 90\%}
\resizebox{\columnwidth}{!}{%
\begin{tabular}{c l l } 
\toprule
Distance \& Angle & Top Class (Confid.) & Second Class (Confid.) \\
\toprule
2' 0\degree & Phone (0.78) & Microwave (0.03)\\
2' 15\degree & Phone (0.60) & Microwave (0.11)  \\
5' 0\degree & Phone (0.71) & Microwave (0.07) \\
5' 15\degree & Phone (0.53) &  Microwave (0.25)  \\
7' 0\degree & Phone (0.47) & Microwave (0.26)  \\
7' 15\degree & Phone (0.59) &Microwave (0.18)     \\
10' 0\degree & Phone (0.70) & Microwave (0.09)  \\
10' 15\degree & Phone (0.43) &Microwave (0.28)     \\
15' 0\degree &  \textcolor{red}{Microwave (0.36)} & Phone (0.20)  \\
20' 0\degree & Phone (0.31) &Microwave (0.10)     \\

\bottomrule
\end{tabular}
}
\label{tab:microwave}
\end{small}

\end{table}

\begin{table}[tb!]
\begin{small}
\centering
\caption{Sticker perturbation attack on the Inception-v3 classifier. The original classification is coffee mug and the attacker's target is cash machine. See example images in Table~\ref{tab:mug-images}. Our targeted-attack success rate is 71.4\%.}
\resizebox{\columnwidth}{!}{%
\begin{tabular}{c l l } 
\toprule
Distance \& Angle & Top Class (Confid.) & Second Class (Confid.) \\
\toprule
8'' 0\degree & Cash Machine (0.53) & Pitcher (0.33)\\
8'' 15\degree & Cash Machine (0.94) & Vase (0.04)  \\
12'' 0\degree & Cash Machine (0.66) & Pitcher (0.25) \\
12'' 15\degree & Cash Machine (0.99) &  Vase (<0.01) \\
16'' 0\degree & Cash Machine (0.62) & Pitcher (0.28)  \\
16'' 15\degree & Cash Machine (0.94) & Vase (0.01)     \\
20'' 0\degree & Cash Machine (0.84) & Pitcher (0.09)  \\
20'' 15\degree & Cash Machine (0.42) & Pitcher (0.38)     \\
24'' 0\degree &  Cash Machine (0.70) & Pitcher (0.20)  \\
24'' 15\degree & \textcolor{red}{Pitcher (0.38)} & Water Jug (0.18)     \\
28'' 0\degree & \textcolor{red}{Pitcher (0.59)} & Cash Machine (0.09)     \\
28'' 15\degree & Cash Machine (0.23) & Pitcher (0.20)     \\
32'' 0\degree & \textcolor{red}{Pitcher (0.50)} & Cash Machine (0.15)     \\
32'' 15\degree & \textcolor{red}{Pitcher (0.27)} & Mug (0.14)     \\
\bottomrule
\end{tabular}
}
\label{tab:mug}
\end{small}

\end{table}

\begin{table}[tb!]
    \small
    \centering
    \caption{Uncropped images of the microwave with an adversarial sticker designed for Inception-v3.}
    \resizebox{\columnwidth}{!}{%
    \begin{tabular}{cM{25mm}M{25mm}M{25mm}M{25mm}}
    \toprule
    Distance/Angle & Image & Distance/Angle & Image   \\
   \toprule
    2\texttt{'}0\degree & \includegraphics[width=0.25\columnwidth]{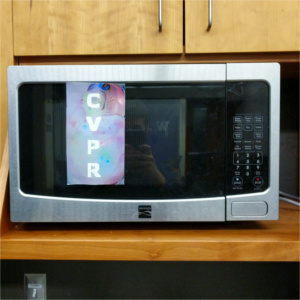} &
    2\texttt{'}15\degree & \includegraphics[width=0.25\columnwidth]{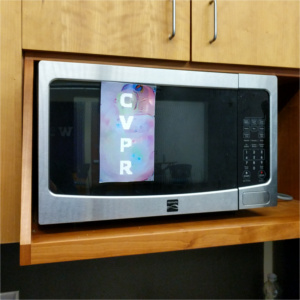} \\
    
    5\texttt{'}0\degree & \includegraphics[width=0.25\columnwidth]{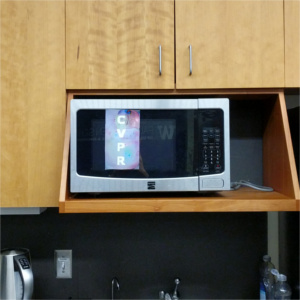} & 
    5\texttt{'}15\degree & \includegraphics[width=0.25\columnwidth]{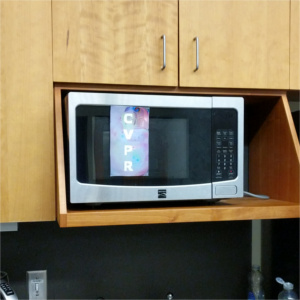} \\
    
    7\texttt{'}0\degree & \includegraphics[width=0.25\columnwidth]{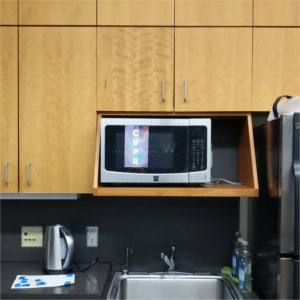} & 
    7\texttt{'}15\degree & \includegraphics[width=0.25\columnwidth]{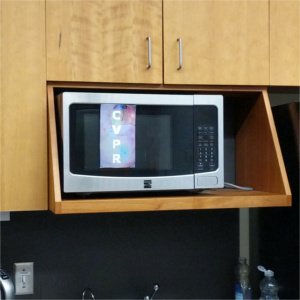}   \\
    
    10\texttt{'}0\degree & \includegraphics[width=0.25\columnwidth]{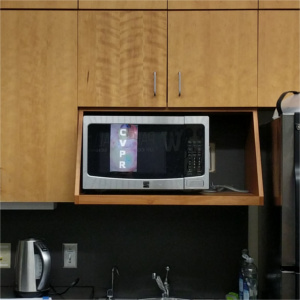} &
    10\texttt{'}15\degree & \includegraphics[width=0.25\columnwidth]{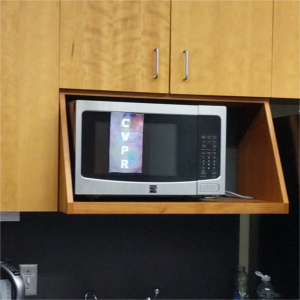} \\
    
    15\texttt{'}0\degree & \includegraphics[width=0.25\columnwidth]{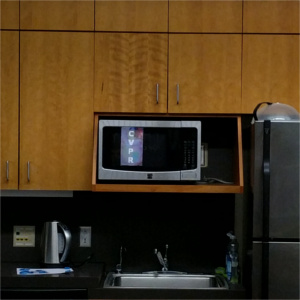} &
    20\texttt{'}0\degree & \includegraphics[width=0.25\columnwidth]{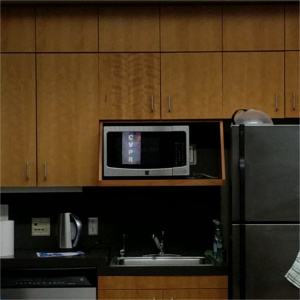} \\

    \\ \bottomrule
    \end{tabular}
    }

\label{tab:microwave-images}
\end{table}

\begin{table}[tb!]
    \small
    \centering
    \caption{Cropped Images of the coffee mug with an adversarial sticker designed for Inception-v3.}
    \resizebox{\columnwidth}{!}{%
    \begin{tabular}{cM{25mm}M{25mm}M{25mm}M{25mm}}
    \toprule
    Distance/Angle & Image & Distance/Angle & Image   \\
   \toprule
    8\texttt{''}0\degree & \includegraphics[width=0.25\columnwidth]{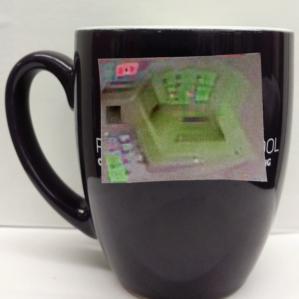} &
    8\texttt{''}15\degree & \includegraphics[width=0.25\columnwidth]{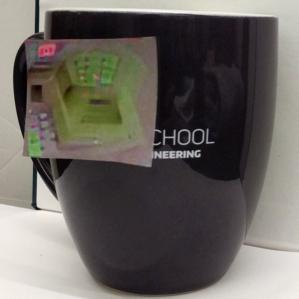} \\
    
    12\texttt{''}0\degree & \includegraphics[width=0.25\columnwidth]{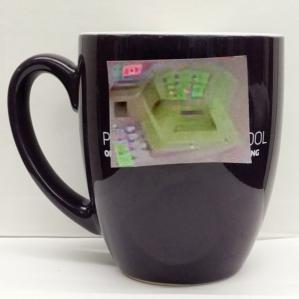} &    12\texttt{''}15\degree & \includegraphics[width=0.25\columnwidth]{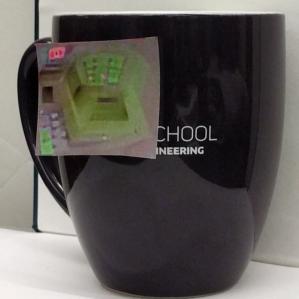} \\
    
    16\texttt{''}0\degree & \includegraphics[width=0.25\columnwidth]{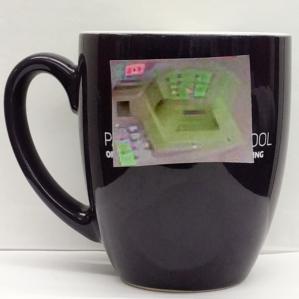} &
    16\texttt{''}15\degree & \includegraphics[width=0.25\columnwidth]{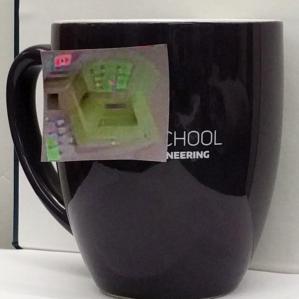} \\
    
    20\texttt{''}0\degree & \includegraphics[width=0.25\columnwidth]{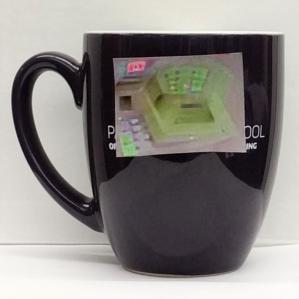} &
    20\texttt{''}15\degree & \includegraphics[width=0.25\columnwidth]{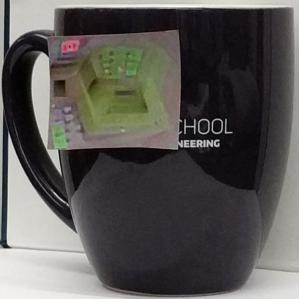}\\
    
    24\texttt{''}0\degree & \includegraphics[width=0.25\columnwidth]{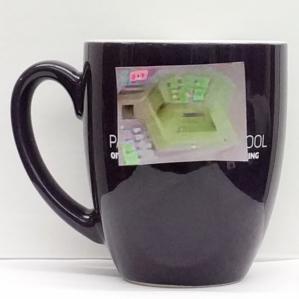} &    24\texttt{''}15\degree & \includegraphics[width=0.25\columnwidth]{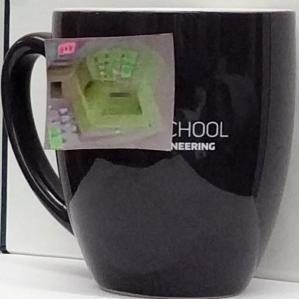} \\

    28\texttt{''}0\degree & \includegraphics[width=0.25\columnwidth]{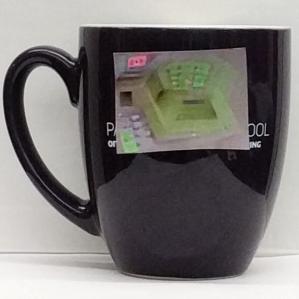} &        28\texttt{''}15\degree & \includegraphics[width=0.25\columnwidth]{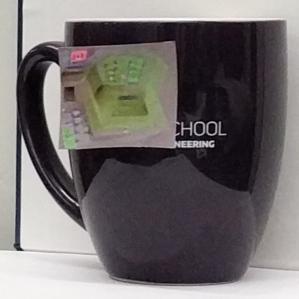} \\

    32\texttt{''}0\degree & \includegraphics[width=0.25\columnwidth]{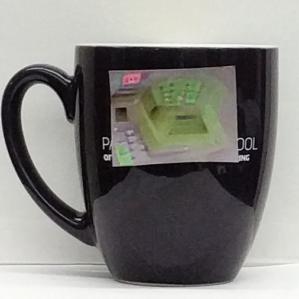} &        32\texttt{''}15\degree & \includegraphics[width=0.25\columnwidth]{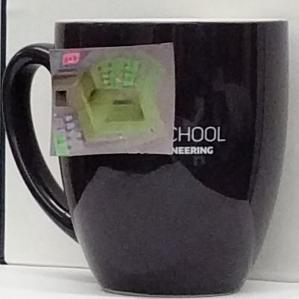} \\

    \\ \bottomrule
    \end{tabular}
    }

\label{tab:mug-images}
\end{table}

\eat{
\begin{figure}
  \centering
  \includegraphics[width=0.47\columnwidth]{new_inception_figs/0.png}
  \includegraphics[width=0.49\columnwidth]{new_inception_figs/2feet.png}
  
    \caption{Physical adversarial example against the Inception-v3 classifier. The left shows the original cropped image identified as microwave (85.2\%) while the right shows the cropped physical adversarial example identified as phone (77.8\%).}
    \label{fig:inception}
\end{figure}}

\section{Discussion}

\noindent\textbf{Black-Box Attacks.} Given access to the target classifier's network architecture and model weights, \algshort can generate a variety of robust physical perturbations that fool the classifier. Through studying a white-box attack like \algshort, we can analyze the requirements for a successful attack using the strongest attacker model and better inform future defenses. Evaluating \algshort in a black-box setting is an open question.

\noindent\textbf{Image Cropping and Attacking Detectors.} When evaluating \algshort, we manually controlled the cropping of each image every time before classification. This was done so the adversarial images would match the clean sign images provided to \algshort. Later, we evaluated the camouflage art attack using a pseudo-random crop with the guarantee that at least most of the sign was in the image. Against LISA-CNN, we observed an average targeted attack rate of 70\% and untargeted  attack rate of 90\%. Against GTSRB-CNN, we observed an average targeted attack rate of 60\% and untargeted attack rate of 100\%. We include the untargeted attack success rates because causing the classifier to not output the correct traffic sign label is still a safety risk. Although image cropping has some effect on the targeted attack success rate, our recent work shows that an improved version of \algshort can successfully attack object detectors, where cropping is not needed~\cite{yoloblog}.


\section{Conclusion}
We introduced an algorithm (\algshort) that generates robust, physically realizable adversarial perturbations. Using \algshort, and a two-stage experimental design consisting of lab and drive-by tests, we contribute to understanding the space of physical adversarial examples when the \textit{objects themselves} are physically perturbed. We target road-sign classification because of its importance in safety, and the naturally noisy environment of road signs. Our work shows that it is possible to generate physical adversarial examples robust to widely varying distances/angles. This implies that future defenses should not rely on physical sources of noise as protection against physical adversarial examples.  

\noindent\textbf{Acknowledgements.}
We thank the reviewers for their insightful feedback. This work was supported in part by NSF grants 1422211, 1616575, 1646392, 1740897, 1565252, Berkeley Deep Drive, the Center for Long-Term Cybersecurity, FORCES (which receives support from the NSF), the Hewlett Foundation, the MacArthur Foundation, a UM-SJTU grant, and the UW Tech Policy Lab.

{\small
\bibliographystyle{ieee}
\bibliography{reference}
}

\end{document}